\documentclass[aps,prl,twocolumn,superscriptaddress,showpacs]{revtex4-1}
\usepackage{graphicx}
\usepackage{amsmath,amssymb}

\begin{document}
%-------------------------------------------------------------------------------------------
\title{Fast and Quasideterministic Single Ion Source from a Dipole-Blockaded Atomic Ensemble}

\author{C. Ates}
\author{I. Lesanovsky}
\affiliation{School of Physics and Astronomy, University of Nottingham, Nottingham, NG7 2RD, United Kingdom}
\author{C. S. Adams}
\author{K. J. Weatherill}
\affiliation{Joint Quantum Centre (JQC) Durham-Newcastle, Department of Physics, Durham University, Durham, DH1 3LE, United Kingdom}
\date{\today}

\begin{abstract}
We present a fast and quasideterministic protocol for the production of single ions and electrons from a cloud of laser-cooled atoms. The approach is based on a two-step process where first a single Rydberg atom is photoexcited from a dipole-blockade configuration and subsequently ionized by an electric field pulse. We theoretically describe these excitation-ionization cycles via dynamical quantum maps and observe a rich behavior of the ionization dynamics as a function of laser Rabi frequency, pulse duration and particle number. Our results show that a fast sequential heralded production of single charged particles is achievable even from an unstructured and fluctuating atomic ensemble.
\end{abstract}


\maketitle

The ability to place single ions into a medium or onto a surface with high precision opens up exciting possibilities for new types of nanofabricated devices and processes in materials science  \cite{shin05,lans08,review,donk10,fuec12}. Techniques such as scanning tunneling microscopy \cite{ibm} and focused ion beam single ion implantation \cite{FIBSII} have been very successful in accurate placing of single ions. However, scanning tunneling microscopy is relatively slow and focused ion beam single ion implantation is stochastic and therefore governed by Poisson statistics. Ideally one would like fast, precise and fully deterministic single ion delivery. A promising route towards this goal is to use laser cooling and trapping techniques \cite{adams1997}, which allow exquisite control over neutral atoms, therefore enabling "atom optics" to become a realistic prospect for nanotechnology applications \cite{rohw07}. By photoionizing ultracold atoms one can then transfer the precision control of neutral atoms onto charged particle sources. This approach has sparked research into the development of a new generation of monochromatic ion and electron beams \cite{hans2008,taba10,mccu11}. Because these "cold" beams originate from an extended source and are demagnified by a factor of several thousand, the ions can be delivered with nanoscale precision and are less sensitive to vibrations and instabilities than pointlike ion sources \cite{stee11}

\begin{figure}
\centering
\includegraphics[width=0.95\columnwidth]{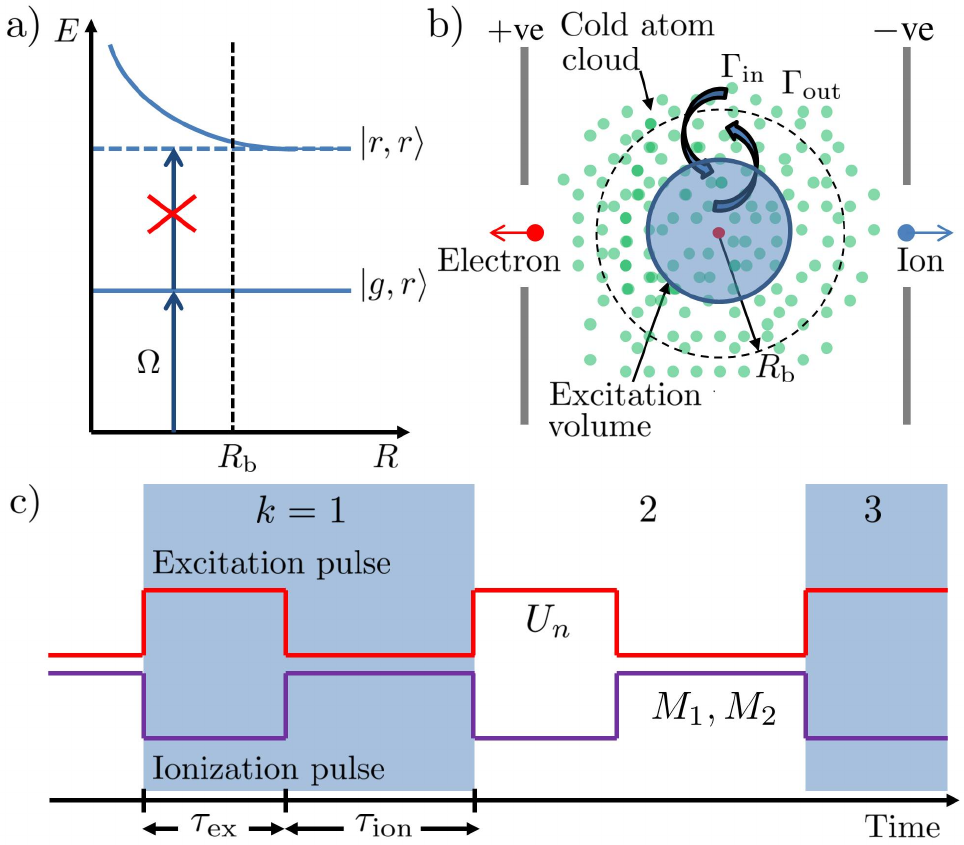}
\caption{a) Dipole blockade. Energy level scheme of a pair of atoms that are resonantly laser-excited from the ground state $|g\rangle$ to a Rydberg state $|r\rangle$ with Rabi frequency $\Omega$. Because of strong interactions between Rydberg atoms the pair state $|r,r\rangle$ experiences a distance-dependent energy shift. Below a critical distance $R_{\rm b}$ - the blockade radius - this shift becomes larger than the Rabi frequency. Here $|r,r\rangle$ becomes energetically inaccessible and only a single Rydberg atoms is excited. b) Schematic of the ion source. Ultracold atoms are excited within the shaded volume through appropriately focussed laser beams. The extension of the excitation volume is smaller than the blockade radius, $R_{\rm b}$. Atoms can diffuse in and out of the region with rate $\Gamma_{\rm in}$ and $\Gamma_{\rm out}$ respectively. An electric field is applied ionizing the Rydberg atom and a single ion and electron are emitted in opposite directions. The field-generating electrodes (carrying charges $\pm v e$) are indicated at the right- and left-hand side of the atomic cloud.}
\label{fig:setup}
\end{figure}

Thus far there have been two main approaches to making single ion sources based upon laser cooling. First, laser-cooled ions held in a trap can be ejected on demand. For example, linear Paul traps can hold strings of ions which can be emitted deterministically \cite{schn2009,izaw10}. Another approach is to trap a single atom, for example in a magneto-optical trap (MOT)  \cite{hill03} or optical trap \cite{henk10}, and then photoionize it. While these approaches show great promise, the trap loading remains random and at present the repetition rate is slow (of the order Hertz).

In this letter we propose an approach that bypasses the relatively slow process of carefully preparing an initial state. This enables the fast and quasideterministic sequential production of ions out of a standard sample of laser-cooled neutral atoms. By analyzing in detail the quantum dynamics of the ion source we identify various dynamical regimes and discuss the temporal ion emission statistics as a function of experimental parameters and atom diffusion. We show that the system not only is suited for technological applications but also provides an interesting platform for the study of nonequilibrium dynamics.

The protocol underlying the envisioned ion source is based on a two-step process: At first, a tightly focused laser beam that is resonant with the transition to a high-lying Rydberg state is irradiated onto an ensemble of ground state atoms. The use of Rydberg states ensures that at most one ion is created in the subsequent step, which consists of an electric field pulse ionizing the Rydberg atom. The reason for the excitation of only a single atom is the dipole blockade \cite{jaksch00,lukin01,urban09,gaetan09,dudin12} which is a consequence of strong and long-range dispersive forces acting among atoms in Rydberg states. These lead to energy level shifts that prevent the laser excitation of many-body states containing more than one Rydberg excitation, cf.\ Fig.\ \ref{fig:setup}(a). An experimental setup depicting the above scheme is sketched in Fig.\ \ref{fig:setup}(b). Here, an ensemble of ultracold atoms is prepared between two electrodes. The excitation lasers are focused such that the overlap of the beams creates an \emph{excitation volume} with a characteristic size that is smaller than the blockade radius $R_{\rm b}$. In a steady state MOT, this can, for instance be achieved by using a three-step excitation scheme \cite{carr12} thereby avoiding avalanche ionization resulting  from excitation over an extended region \cite{robe13}.  We assume that the Rabi frequency is uniform over the excitation region, which can be achieved using flattop beams \cite{reet08}. The application of a small electric field results in the field ionization of the excited atom and the resulting ion and its concomitant electron are ejected from the ensemble in opposite directions as shown in the Fig.~1 (b). The lighter electron can then be detected to "herald" the creation of the ion.

In what follows we describe a simple model that captures the essential dynamical processes of such an ion source and which permits a detailed study of the statistics of the ion emission. An excitation-ionization cycle consists of a resonant laser pulse of duration $\tau_{\text{ex}}$ followed by an electric field pulse of duration $\tau_{\text{ion}}$. The Hamiltonian describing the excitation step is given by $H = \frac{\Omega}{2} \left( a^{\dagger} b + b^{\dagger} a \right)$, where $\Omega$ denotes the single-atom Rabi-frequency, $a^{\dagger}$ ($a$) a bosonic operator that creates (annihilates) a particle in the electronic ground state and $b^{\dagger}$ ($b$) an operator that creates (annihilates) a Rydberg atom. In order to take the Rydberg blockade into account in a simple way we impose fermionic anti-commutator relations $\{ b,b^{\dagger} \} =1$ and $\{ b,b\}= \{b^{\dagger},b^{\dagger}\} =0$ on the $b$ operators, so that $b^2 =0$. This accounts for the fact that a double Rydberg excitation within the blockade volume is impossible \cite{footnote}.

The dynamics of the incoherent ionization step is governed by a Markovian Master equation for the density matrix $\rho$ of the (unionized) atoms contained in the excitation volume. In order to explicitly resolve the number of ions we project $\rho$ onto the subspace in which $m \ge 0$ ions have already been created. If $\rho^{(m)}$ denotes the projected density matrix then $\rho = \sum_m \rho^{(m)}$ and the master equation governing its dynamics is \cite{zoll87}
\begin{equation}
i \partial_t \rho^{(m)} = \gamma \left[ b \rho^{(m-1)} b^{\dagger} \left( 1- \delta_{m,0} \right) - \frac{1}{2} \left\{ b^{\dagger} b, \rho^{(m)}  \right\} \right]
\label{eq:master}
\end{equation}
with ionization rate $\gamma$.

Since coherent laser-excitation and incoherent ionization happen in sequence, we can efficiently describe the dynamics of a whole excitation-ionization cycle within the framework of dynamical quantum maps \cite{haro06}. The time evolution during the $k$-th excitation-ionization cycle is given as follows: Starting at time $t_{k-1}$ the laser pulse coherently excites an atom in the excitation volume, i.e., the density matrix $\rho_{k-1}$ undergoes a unitary evolution $\rho^{\prime} = U\rho_{k-1}U^{\dagger}$ with $U(\tau_{\text{ex}}) = \exp(-i H \tau_{\text{ex}})$. Since the laser-excitation does not change the number of particles in the volume, $U$ is block diagonal in particle number space and the coherent dynamics for an ensemble of $n$ atoms takes place in the subspace spanned by the states $|g\rangle^{\otimes n}$ and $\mathcal{S}\left[|g\rangle^{\otimes (n-1)}\otimes|r\rangle\right]/\sqrt{n}$, where $\mathcal{S}$ is the symmetrization operator. The corresponding time evolution operator for the $n$ particle block is
\begin{equation}
\label{eq:unitary}
U_n = \left( \begin{array}{cc}
\cos \frac{\sqrt{n} \Omega}{2} \tau_{\text{ex}} & -i \sin \frac{\sqrt{n} \Omega}{2} \tau_{\text{ex}} \\
-i \sin \frac{\sqrt{n} \Omega}{2}  \tau_{\text{ex}} & \cos \frac{\sqrt{n} \Omega}{2} \tau_{\text{ex}}
\end{array} \right) ,
\end{equation}
showing explicitly the $\sqrt{n}$ enhancement of the atom-field coupling due to the blockade effect \cite{dudin12}. The subsequent ionization pulse maps the density matrix $\rho^{\prime}$ to $\rho_k = M_1\rho^{\prime} M_1^{\dagger} + M_2\rho^{\prime} M_2^{\dagger}$, where
\begin{equation}
\label{eq:Kraus}
M_1 = \left( \begin{array}{cc}
1 & 0 \\
0 & e^{-\gamma \tau_{\text{ion}} /2}
\end{array} \right),\,
M_2 = \left( \begin{array}{cc}
0 & \sqrt{ 1 - e^{-\gamma \tau_{\text{ion}} }} \\
0 & 0
\end{array} \right)
\end{equation}
are Kraus operators. $M_1$ describes the dynamics induced by the anti-commutator term of the master equation (\ref{eq:master}), while $M_2$ captures the quantum jump processes induced by the first term on the right hand side of eq.\ (\ref{eq:master}).

The density matrix of the subspace, where exactly $m$ ions have been produced after the $k$-th excitation-ionization cycle can now be obtained iteratively,
\begin{eqnarray}
\label{eq:iteration}
\rho^{(0)}_k &=& \mathcal{K}^{(0)}_1 \rho^{(0)}_{k-1} \\
\rho^{(m)}_k &=& \left[ \mathcal{K}^{(m)}_1 \rho^{(m)}_{k-1} \left(1-\delta_{km} \right)+  \mathcal{K}^{(m)}_2 \rho^{(m-1)}_{k-1} \right] \Theta(k-m) \nonumber
\end{eqnarray}
with $\mathcal{K}^{(m)}_1 \rho^{(m)}_{k-1} = M_1 U_{N-m} \; \rho^{(m)}_{k-1} \; U_{N-m}^{\dagger} M_1^{\dagger}$ and $\mathcal{K}^{(m)}_2 \rho^{(m-1)}_{k-1} = M_2 U_{N-m+1} \; \rho^{(m-1)}_{k-1} \; U_{N-m+1}^{\dagger} M_2^{\dagger}$. Here, $N$ denotes the number of atoms in the focal volume at $t=0$ and $\Theta(x)$ is the Heaviside step function.

Let us now apply this theoretical framework to analyze the dynamics of the ion source in different operational regimes:

\begin{figure}
\centering
\includegraphics[width=0.9\columnwidth]{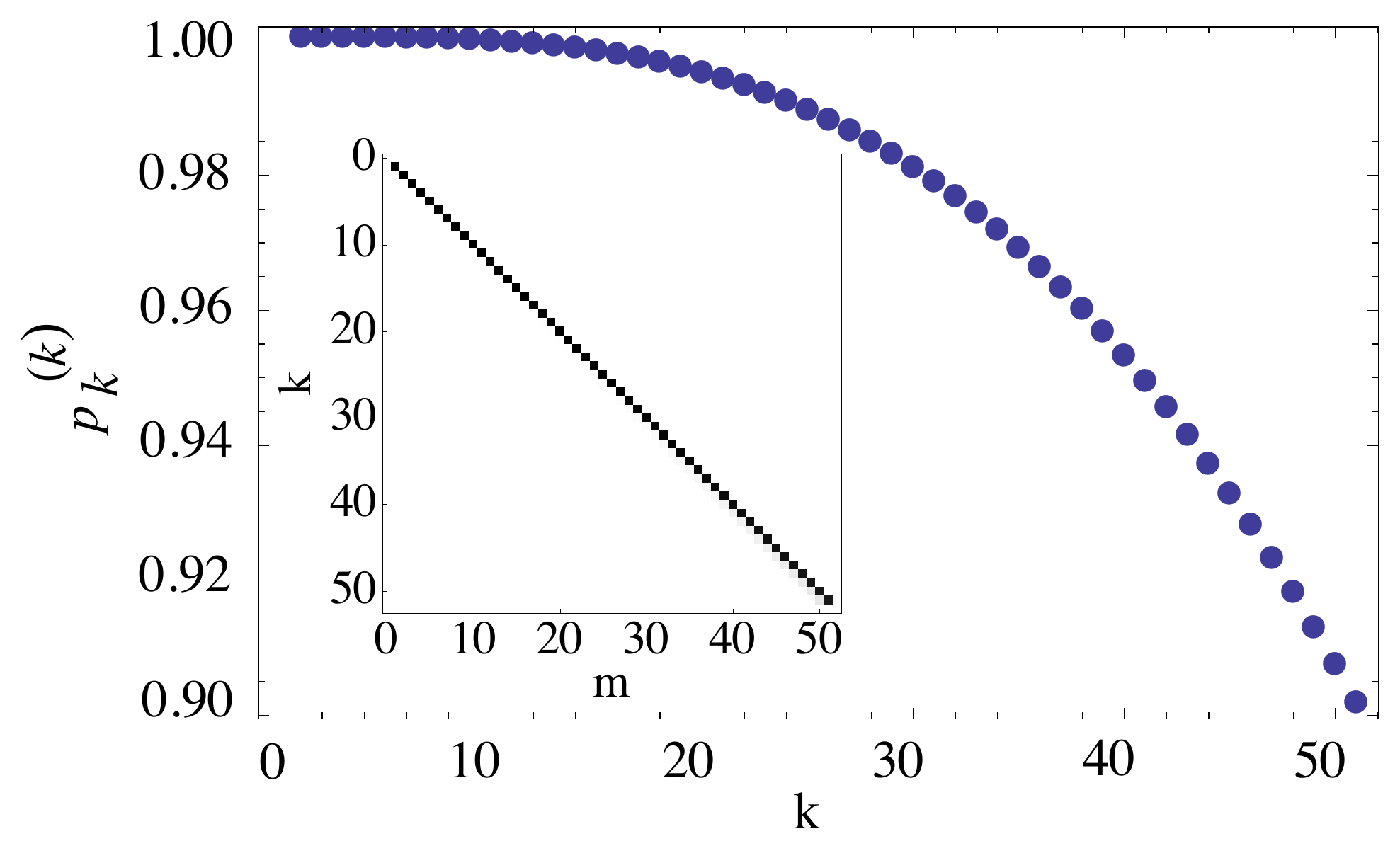}
\caption{Quasideterministic ion production. Probability $p^{(k)}_k$ of having produced $k$ ions after $k$ cycles for initial particle number $N=500$ and excitation time $\tau_{\text{ex}} = \pi /(\sqrt{N} \Omega)$ (the other parameters are: Rabi frequency $\Omega = 4 \gamma$ ionization pulse length $\tau_{\text{ion}} = 16 \gamma^{-1}$). Inset: Probability distribution $p^{(m)}_k$ of having produced $m$ ions after $k$ cycles. The color map is chosen such that white (black) corresponds to 0 (1). The data show that a single ion is created with nearly unit probability after each excitation-ionization cycle.} \label{fig:repeat}
\end{figure}
\noindent\textit{(i) Quasi-deterministic regime in a microtrap} - Here we consider an atomic ensemble which is confined within a trap volume which is much smaller than the blockade radius. We start initially from a fixed number of trapped atoms $N$ and the successive application of the excitation-ionization cycle will lead eventually to a depletion of the trap. For our numerical example we consider an initial atom number of $N=500$ - a situation which can be achieved, for example, in the case of an atomic Bose-Einstein condensate held in a tight optical trap. The Rydberg excitation is carried out with a $\pi$-pulse, with respect to the collective Rabi frequency $\sqrt{N} \Omega$ \cite{heid07}. In the regime of a relatively large initial number of atoms, i.e. $N \gg 1$, this collective Rabi frequency changes negligibly with the emission of each ion. Thus, each excitation cycle of length $\tau_{\text{ex}}=\pi/\sqrt{N}\Omega$ will - with probability close to unity -  excite a single Rydberg atom, and this excitation process will remain efficient even after many ions have been emitted. This is indeed illustrated in Fig. \ref{fig:repeat} which shows the probability of having ionized exactly $k$ atoms after $k$ excitation-ionization cycles. In addition, the inset shows the probability distribution $p^{(m)}_k$ of having ionized $m$ atoms after $k$ cycles. Hence, for a fully deterministic single ion source this quantity is 1 for $m=k$ and 0 elsewhere. The data indeed reflects that a single ion is produced in each cycle until the depletion of the trap begins to alter the collective Rabi frequency significantly. In principle the Rabi frequency could be adapted dynamically in order to account for atom loss. However, we can see that even in the static case, many atoms can be rapidly extracted from the sample with high probability. Using typical experimental parameters, i.e. an excitation time $\tau_\mathrm{ex}\sim 10$ ns and an ionization rate $\gamma\sim 100$ kHz an ion production rate up to $100$ kHz appears feasible.

\begin{figure}
\centering
\includegraphics[width=0.95\columnwidth]{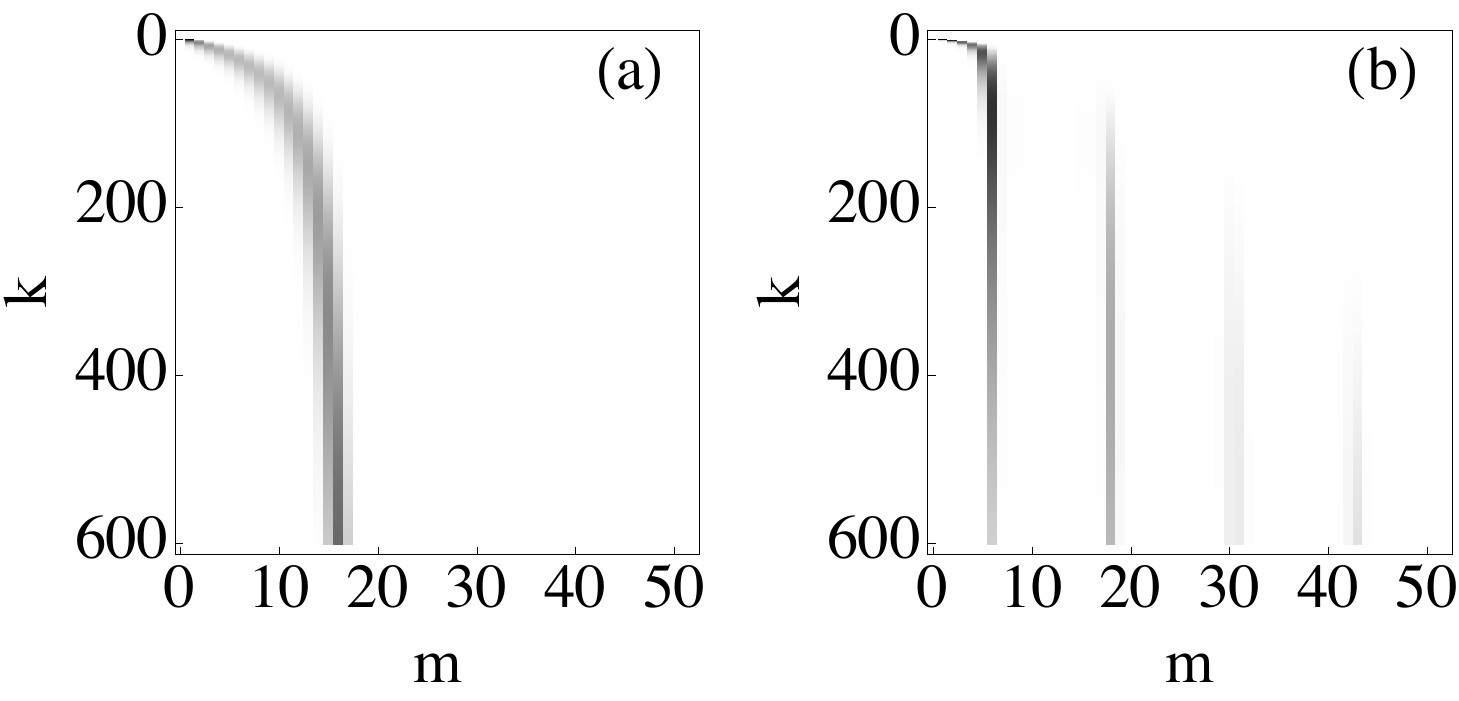}
\caption{Trapped states. Probability distribution $p^{(m)}_k$ of having produced $m$ ions after $k$ excitation-ionization cycles for a fully blockaded sample initially prepared with $N=500$ atoms. (a) The Rydberg excitation is performed using a single-atom $\pi$-pulse, i.e., $\tau_{\text{ex}} = \pi/\Omega$. (b) The Rydberg excitation pulse length is $\tau_{\text{ex}} = 7.1 \pi/\Omega$. The values of the other parameters are $\gamma=4\Omega$ and $\tau_{\text{ion}} = 4 \Omega^{-1}$. The colormap is chosen such that white (black) corresponds to 0 (1).} \label{fig:trapped}
\end{figure}
\noindent\textit{(ii) Trapped states in a microtrap} - We are now interested in the ion emission dynamics for the case when the excitation pulse length is chosen independent of the particle number in the excitation volume, i.e, without taking the collective laser-coupling induced by the dipole blockade into account.
Let us first consider the situation in which Rydberg atoms are excited with a single-atom $\pi$-pulse, i.e. $\tau_{\text{ex}}=\pi\Omega^{-1}$. The evolution of the corresponding probability density is shown in Fig. \ref{fig:trapped}(a). We find that initially the mean number of emitted ions $m$ grows with the number of cycles $k$. However, in this example with $500$ initial atoms, the probability of emitting ions beyond the $250^{th}$ cycle is strongly suppressed. In fact the data shows that no more than $16$ ions will be emitted in total even if the number of cycles is increased. The reason for that is the emergence of trapped states which occur when $\sqrt{N-m}\Omega\tau_\mathrm{ex}/2$ becomes an integer multiple of $\pi$. Here the time evolution operator (\ref{eq:unitary}) becomes the identity (up to a phase factor of $\pm 1$) and consequently the excitation step does not generate Rydberg atoms. This effect is analogous to the occurrence of trapped states in micromasers. These are photonic number states such that atoms transiting the cavity undergo a $2\pi$ pulse (or an integer multiple of it) \cite{fil86,weid99}. A further interesting dynamical regime is reached when the Rydberg excitation pulse is close to a $\nu\, \pi$-pulse with $\nu=1,3,5,...$. In this case quasi-trapped states can occur that have a striking effect on the ionization dynamics. The corresponding probability distribution for an excitation pulse length of $\tau_{\text{ex}} = 7.1 \pi/\Omega$ is depicted in Fig.\ \ref{fig:trapped}(b). As in the previous example a trapped state is reached after a few initial cycles. However, since $\sqrt{n}\Omega\tau_\mathrm{ex}/2$ can never be a strict multiple of $\pi$ the trapping is not perfect and we observe a slow leakage into other quasi-trapped states. Eventually, this leads to a multi-modal ion distribution that peaks at ion numbers $m$ where $\sqrt{N-m}\,\Omega\tau_\mathrm{ex}/2$ is close to a multiple integer of $\pi$. This shows that in order to avoid trapping effects and to use the proposed device as single-ion source the excitation pulse should be chosen to be a $\pi$-pulse with respect to the  collective Rabi frequency.

\begin{figure}
\centering
\includegraphics[width=0.95\columnwidth]{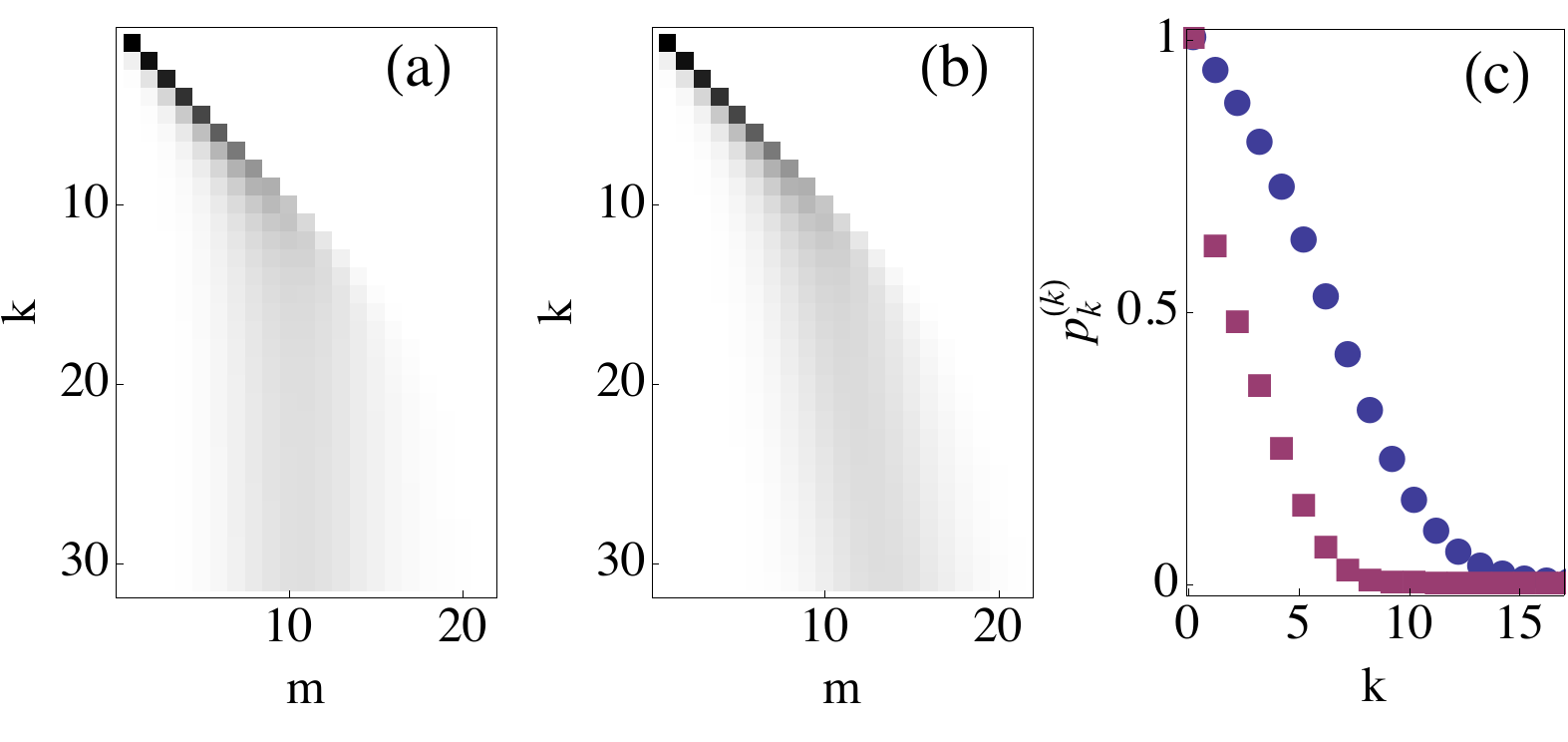}
\caption{Ion creation from an extended gas cloud. Probability $p^{(m)}_k$ of having produced $m$ ions after $k$ excitation-ionization cycles (a) without particle diffusion and (b) for diffusion rate $\kappa=10^{-3} \gamma$. The initial number of particles in the excitation volume is distributed following a Poisson distribution with mean $\overline{n} =10.2$. The data shown are averaged over this initial particle number distribution. The excitation pulse length is $\tau_{\text{ex}} = \pi /{\sqrt{\overline{n}} \Omega}$. The Rabi frequency is $\Omega = 5 \gamma$ and the ionization pulse length is $\tau_{\text{ion}} = 20 \gamma^{-1}$. The colormap is chosen such that white (black) corresponds to 0 (1). (c) Influence of the choice of the excitation pulse length on the emission characteristics of the ion source. Shown is the probability of having produced $k$ ions after $k$ excitation-ionization cycles for $\tau_{\text{ex}} = \pi /{\sqrt{\overline{n}} \Omega}$ (circles) and $\tau_{\text{ex}} = \pi /{ \Omega}$, i.e. a single-atom $\pi$-pulse (squares). All other parameters are those of panel (b).
}\label{fig:cycles}
\end{figure}
\noindent\textit{(iii) Excitation from an extended atomic cloud} - So far we have considered a fully blockaded dense atomic sample in a microtrap with radius much smaller than $R_{\rm b}$. Let us now discuss the situation usually encountered in extended systems like standard MOTs. Here the atom density is relatively small and, moreover, atoms can diffuse in and out of the excitation volume as depicted in Fig. \ref{fig:setup}(b). In order to account for the effects of diffusion we use a simple rate model: Atoms can enter the excitation volume at a constant rate $\Gamma_\mathrm{in}$ and leave it at a rate $\Gamma_\mathrm{out}=\kappa n$, where $n$ is the number of atoms in the volume and $\kappa$ is the diffusion rate. In an equilibrium gas cloud we have $\Gamma_\mathrm{in}=\Gamma_\mathrm{out}$ and hence $\Gamma_\mathrm{in}=\kappa \bar{n}$ where $\bar{n}$ is the mean number of atoms in the excitation volume. This simple model yields a Poissonian atom distribution with mean $\overline{n}$ for finding $n$ atoms in the excitation region.

To describe the influence of diffusion on the ion source dynamics we exploit the fact that the excitation cycle ($\tau_{\text{ex}} \sim 10$ns) is typically much shorter than the ionization step ($\tau_{\text{ion}} \sim 10 \mu$s). Thus we assume that diffusion is only relevant during the latter. This is consistent with the frozen gas approximation \cite{mour98} usually employed in the description of the laser excitation of cold Rydberg gases. Moreover, we consider the limit of slow diffusion, i.e. we assume $\kappa\tau_\mathrm{ion}\gg 1/(2\bar{n})$, which means that during the ionization cycle only one atom will enter/leave the excitation volume at a time. Effectively this amounts to solving the diffusion equation up to first order. Within these approximations we can again use the above-employed theoretical framework of dynamical quantum maps and arrive at dynamical equations analogous to (\ref{eq:iteration}) (see supplementary material).

Let us now investigate the excitation dynamics. We assume that the initial mean number of atoms in the excitation volume is $\bar{n}=10.2$. With a realistic blockade radius of 5~$\mu$m this would correspond to  typical MOT densities of 10$^{10}$ atoms/cm$^3$. In Fig.\ \ref{fig:cycles} (a) and (b) we show the corresponding probability distributions $p^{(m)}_k$ for the case without diffusion and for a diffusion rate of $\kappa=10^{-3} \gamma$, respectively. The data are averaged over the distribution of the initial number of particles in the volume and the excitation pulse length is fixed at $\tau_{\text{ex}} = \pi /{\sqrt{\overline{n}} \Omega}$. This choice therefore corresponds to a $\pi$-pulse with respect to the collective Rabi frequency $\sqrt{\overline{n}} \Omega$. Comparing Figs.\ \ref{fig:cycles}(a) and (b) we see that diffusion leads to a slow drift of the ion number distribution to larger mean values with increasing number of excitation-ionization cycles. This drift stems from the fact that the excitation volume is quickly depleted due the ionization process which is much faster than the diffusion. At long times (large $k$) further ions can only be produced when particles diffuse into the empty excitation volume leading to a slow increase of the total number of emitted ions.

More interestingly, however, for the first few cycles the ion source produces one ion per cycle with a high probability and is therefore quasi-deterministic. Note that this is despite averaging over the initial atom number distribution. Therefore, even when operated in an extended atomic cloud, the behavior of the ion source is very similar to that in a fully blockaded sample discussed in \textit{(i)}. This is further corroborated by Fig.\ \ref{fig:cycles}(c) where we show the probability of having emitted $k$ ions after $k$ excitation-ionization cycles. For comparison we also show the probability for the case in which the Rydberg excitation is performed with a $\pi$-pulse w.r.t. the single atom Rabi frequency $\Omega$. The data shows that the performance of the ion source is indeed significantly improved when the excitation pulse is chosen as $\pi$-pulse w.r.t. the collective Rabi frequency $\sqrt{\bar{n}}\Omega$. Note, that the performance of the ion-source can be enhanced further by resetting the ion source after the few cycles during which quasi-deterministic ion emission is taking place. This could be done by refocusing the excitation laser to a different part of the gas cloud thereby avoiding the depletion of the excitation volume and the broadening of the ion distribution visible in Fig. \ref{fig:cycles}(a,b). The movement of the source can easily be compensated by ion optics to ensure the ion is imaged on to the correct position. The fidelity of the single ion emission can be tested by performing number resolved coincidence measurements \cite{loch12}.

In conclusion, we have presented and analyzed theoretically a fast and heralded single ion source based on a dipole-blockaded atomic ensemble. The source can be operated in a variety of dynamical regimes and permits the quasi-continuous and quasi-deterministic production of single ions even from a extended gas cloud. The statistics of the ion emission shows some interesting dynamical features such as the emergence of `trapped' states, reminiscent of micromaser physics. We expect that the presented scheme can be of practical relevance for  applications in materials science.

\begin{acknowledgments}
\emph{Acknowledgements --- }
I. L. acknowledges funding through EPSRC and the ERA-NET CHIST-ERA (R-ION
consortium) and the Leverhulme Trust. C.A. acknowledges support through a Feodor-Lynen Fellowship of the Alexander von Humboldt Foundation. C.S.A acknowledges funding through EPSRC.
\end{acknowledgments}

%\bibliographystyle{apsrev4-1}
%\bibliography{/Users/cenap/Documents/Literatur/cenap}

\begin{thebibliography}{99}
\bibitem{footnote} The probability $p_2(t)$ of having excited two Rydberg atoms after a time $t$ in the excitation volume can be estimated using time-dependent perturbation theory, $p_2(t) = n(n-1) \Omega^2 \sin^2 (\sqrt{n} \Omega t)/(2 V)$, where $n$ is the number of atoms and $V$ the minimal possible interaction strength of two Rydberg atoms within the excitation volume.
\end{thebibliography}

\newpage

\begin{widetext}

\section{Supplementary Material}

As discussed in the article we assume that diffusion is only relevant during the ionization step. Since we have two processes that can change the number of atoms in the focal volume (ionization and diffusion), we need an additional index to label the density matrix, $\rho^{(n,m)}_{\alpha,\beta}$, since the number of atoms in the excitation volume is no longer solely determined by the number of ions produced. Here, $n$ refers to the number of atoms in the volume, $m=0,1,\dots$ to the number of ionization events that have already taken place and $\alpha , \beta = 0\; (1)$ labels the electronic ground state (Rydberg state) of an atom. The evolution equations during the ionization step read

\begin{subequations}
\begin{eqnarray}
\partial_t \rho^{(n,m)}_{00} &=& 
\gamma \rho^{(n+1,m-1)}_{11}
 -\kappa (\overline{n} + n) \rho_{00}^{(n,m)} + \kappa (n+1) \rho_{00}^{(n+1,m)} + \kappa \overline{n} \rho_{00}^{(n-1,m)} \\
\partial_t \rho^{(n,m)}_{11} &=& 
- \gamma \rho^{(n,m)}_{11} \\
\partial_t \rho^{(n,m)}_{01} &=& 
 - \frac{\kappa + \gamma}{2} \rho^{(n,m)}_{01} \\
\partial_t \rho^{(n,m)}_{10} &=& 
- \frac{\kappa + \gamma}{2} \rho^{(n,m)}_{10} ,
\end{eqnarray}
\label{hybrid}
\end{subequations}
where $\overline{n}$ denotes the mean atom number in the excitation volume at equilibrium (i.e., before starting to operate the ion source), $\gamma$ the ionization rate and $\kappa$ the diffusion rate.

In order to obtain the dynamics of the excitation-ionization cycles we now need six operators: $U_n$, $M_1$ and $M_2$ are those that we have already introduced for the diffusionless situation [eqns. (2) and (3) in the article] and in addition $M_3$, $M_4^{\text{in}}$ and 
$M_4^{\text{out}}$ where
\begin{equation}
M_3(n,\tau) = \left( \begin{array}{cc}
\sqrt{1 - \kappa \tau (\overline{n} + n)} & 0 \\
0 & 1
\end{array} \right)
\end{equation}
and
\begin{equation}
M_4^{\text{in}}(\tau) = 
\left( \begin{array}{cc}
\sqrt{\kappa \tau \overline{n}} & 0 \\ 
0 & 1
\end{array} \right) \quad , \quad
M_4^{\text{out}}(n,\tau) = \left( \begin{array}{cc}
\sqrt{\kappa \tau n} & 0 \\
0 & 1
\end{array} \right) .
\end{equation}
The operator $M_3$ gives the time-evolution of the density matrix that is diagonal in the number of particles $n$ within the excitation volume, while the $M_4$'s describe the quantum jumps due to particle diffusion. The $M_4$ operators change the number of particles in the volume without changing the number of ions $m$, while $M_2$ affects both quantum numbers.  In the following it will be convenient to separate these two  processes. To do so, we define block matrices that capture the dynamics within different $n$-subspaces by introducing a new set of Kraus operators. The block diagonal operators are

\begin{subequations}
\begin{equation}
\mathcal{U} = \left(
\begin{array}{ccccc}
\ddots & 0 & \dots & \\
0 & U_n(\tau) & 0 & \dots & \dots \\
\dots & 0 & U_{n+1}(\tau) & 0 & \dots \\
 &\vdots & 0 & \ddots & \ddots
\end{array}
\right)
\end{equation}
\begin{equation}
\mathcal{M}_1 = \left(
\begin{array}{ccccc}
\ddots & 0 & \dots & \\
0 & M_1(\tau) & 0 & \dots & \dots \\
\dots & 0 & M_1(\tau) & 0 & \dots \\
 &\vdots & 0 & \ddots & \ddots
\end{array}
\right)
\end{equation}
\begin{equation}
\mathcal{M}_3 = \left(
\begin{array}{ccccc}
\ddots & 0 & \dots & \\
0 & M_3(n,\tau) & 0 & \dots & \dots \\
\dots & 0 & M_3(n+1,\tau) & 0 & \dots \\
 &\vdots & 0 & \ddots & \ddots
\end{array}
\right) .
\end{equation}
In contrast the jump processes are described by Kraus operators that have their non-trivial blocks on the sub- and super-diagonal in particle number space
\begin{equation}
\mathcal{M}_2 = \left(
\begin{array}{ccccc}
\ddots & M_2(\tau) & \dots & \\
0 & 0 & M_2(\tau) & \dots & \dots \\
\dots & 0 & 0 & M_2(\tau) & \dots \\
 &\vdots & 0 & \ddots & \ddots
\end{array}
\right)
\end{equation}
\begin{equation}
\mathcal{M}_4^{\text{out}} = \left(
\begin{array}{ccccc}
\ddots & M_4^{\text{out}}(n,\tau) & \dots & \\
0 & 0 & M_4^{\text{out}}(n+1,\tau) & \dots & \dots \\
\dots & 0 & 0 & M_4^{\text{out}}(n+2,\tau) & \dots \\
 &\vdots & 0 & \ddots & \ddots
\end{array}
\right)
\end{equation}
\begin{equation}
\mathcal{M}_4^{\text{in}} = \left(
\begin{array}{ccccc}
\ddots & 0 & \dots & \\
M_4^{\text{in}} (\tau) & 0 & 0 & \dots & \dots \\
\dots & M_4^{\text{in}} (\tau) & 0 & 0 & \dots \\
 &\vdots & M_4^{\text{in}} (\tau) & \ddots & \ddots
\end{array}
\right) .
\end{equation}
\end{subequations}
Since the processes coupling different $n$-subspaces do not generate coherences between $n$ and $n^{\prime}$ the density matrix is block diagonal in $n$
\begin{equation}
\rho^{(m)} (\tau) = \left(
\begin{array}{ccccc}
\ddots & 0 & \dots & \\
0 & \rho^{(n,m)} (\tau) & 0 & \dots & \dots \\
\dots & 0 & \rho^{(n+1,m)} (\tau) & 0 & \dots \\
 &\vdots & 0 & \ddots & \ddots
\end{array}
\right) .
\end{equation}
The time evolution of $\rho^{(m)}$ can now be described equivalently to the diffusionless case
\begin{subequations}
\begin{eqnarray}
\rho^{(0)}_k &=& \mathcal{M}_1(\tau_{\text{ion}}) \, \mathcal{M}_3(\tau_{\text{ion}}) \, \mathcal{U} (\tau_{\text{ex}}) \, \rho^{(0)}_{k-1} \, \mathcal{U}^+ (\tau_{ex}) \, \mathcal{M}^+_3 (\tau_{\text{ion}}) \, \mathcal{M}^+_1 (\tau_{\text{ion}}) \nonumber \\
& & {}+ \mathcal{M}_4^{\text{in}} \tau_{\text{ion}}) \, \mathcal{U} (\tau_{\text{ex}}) \, \rho^{(0)}_{k-1} \, \mathcal{U}^+ (\tau_{ex}) \,\left[ \mathcal{M}_4^{\text{in}} \right]^+ \tau_{\text{ion}}) \nonumber \\
& & {}+ \mathcal{M}_4^{\text{out}} \tau_{\text{ion}}) \, \mathcal{U} (\tau_{\text{ex}}) \, \rho^{(0)}_{k-1} \, \mathcal{U}^+ (\tau_{ex}) \,\left[ \mathcal{M}_4^{\text{out}} \right]^+ \tau_{\text{ion}})  \\
%-------------------------------------------------------------------
\rho^{(m)}_k &=& \bigg\{ \Big[ 
 \mathcal{M}_1(\tau_{\text{ion}}) \, \mathcal{M}_3(\tau_{\text{ion}}) \, \mathcal{U} (\tau_{\text{ex}}) \, \rho^{(m)}_{k-1} \, \mathcal{U}^+ (\tau_{ex}) \, \mathcal{M}^+_3 (\tau_{\text{ion}}) \, \mathcal{M}^+_1 (\tau_{\text{ion}}) \nonumber \\
& & {}+ \mathcal{M}_4^{\text{in}} \tau_{\text{ion}}) \, \mathcal{U} (\tau_{\text{ex}}) \, \rho^{(m)}_{k-1} \, \mathcal{U}^+ (\tau_{ex}) \,\left[ \mathcal{M}_4^{\text{in}} \right]^+ \tau_{\text{ion}}) \nonumber \\
& & {}+ \mathcal{M}_4^{\text{out}} \tau_{\text{ion}}) \, \mathcal{U} (\tau_{\text{ex}}) \, \rho^{(m)}_{k-1} \, \mathcal{U}^+ (\tau_{ex}) \,\left[ \mathcal{M}_4^{\text{out}} \right]^+ \tau_{\text{ion}})
\Big](1 - \delta_{km} )+ \nonumber \\
 & & {}+ \mathcal{M}_2(\tau_{\text{ion}}) \, \mathcal{U} (\tau_{\text{ex}}) \, \rho^{(m-1)}_{k-1} \, \mathcal{U}^+ (\tau_{ex}) \, \mathcal{M}^+_2 (\tau_{\text{ion}}) \bigg\} \, \Theta (k-m)
\end{eqnarray}
\end{subequations}

These equations are the generalization of eqn.\ (4) in the article for the case that particles can diffuse in and out of the focal volume. The diffusion process is captured by the Kraus operators $\mathcal{M}_3$, $\mathcal{M}_4^{\text{in}}$ and $\mathcal{M}_4^{\text{out}}$. The operator $\mathcal{M}_3$ captures the damping of the electronic ground state population as well as the electronic coherences between ground and Rydberg state for a given particle number $n$ in the focal volume due to diffusion. The two $\mathcal{M}_4$ operators describe the (incoherent) change in particle number due to diffusion, which is treated as quantum jump processes.    

\end{widetext}

\end{document}